# DEVIATIONS FROM HIERARCHICAL CLUSTERING IN REAL AND REDSHIFT SPACE

Sebastiano Ghigna & Silvio A. Bonometto[1,2],

Luigi Guzzo[3], Riccardo Giovanelli & Martha P. Haynes[4],

Anatoly Klypin [5], Joel R. Primack[6].

[1] Dipartimento di Fisica dell'Università di Milano Via Celoria 16, I-20133 Milano, Italy

[2] INFN, Sezione di Milano,

[3] Osservatorio Astronomico di Brera, Sede di Merate, Via Bianchi 46, I-22055 Merate (CO), Italy

[4] Department of Astronomy and National Astronomy and Ionosphere Center (The National Astronomy and Ionosphere Center is operated by Cornell University under a cooperative agreement with the National Science Foundation) Space Sciences Building, Cornell University, Ithaca, NY 14853, USA

[5] Astronomy Department, New Mexico State University Box 30001 – Dept. 4500, Las Cruces, NM 88003-0001, USA and Astro-Space Center, Lebedev Phys. Inst., Moscow, Russia

[6] Institute for Particle Physics, University of California, Santa Cruz, CA 95064, USA

# Abstract

We discuss the effects of redshift–space distortions on the estimate of high order galaxy correlation functions. We use both the Perseus–Pisces redshift survey and the results of high–resolution N–body simulations to explore the consequences of working in redshift space on the detection of deviations from hierarchical clustering. Both for real and simulated data, significant deviations from hierarchical clustering seem to be present in real space. Their behaviour is coherent with that expected from an initially biased galaxy field, once the displacement from the sites where galaxies formed, due to the nonlinear gravitational evolution, is taken into account. The passage to redshift space has the net effect of filtering out the higher powers required to fit the distribution in real space. This magnifies the distortions operated by nonlinear evolution on the initial distribution, erasing the residual narrow scale range (2–5 $h^{-1}$Mpc) where deviations from hierarchical clustering are still detectable in real space. We conclude that such deviations can hardly be estimated using data in redshift space. Suitable improvements to the techniques of real–space reconstruction should therefore be studied, in order to preserve actual deviations from hierarchical clustering and make use of their possible statistical measures.

# 1. Introduction

The observed large scale (LS) mass distribution is well known to be non-Gaussian. A detailed analysis of $n$-point ($n > 2$) correlation functions aims to discriminate among different kinds of non-Gaussian processes.

In this note we report the results of an analysis of $n$-point ($n = 3, 4$) functions based on neighbour moment evaluation. We compare a volume limited sample (VLS) of real galaxies, extracted from the Perseus–Pisces redshift Survey (PPS), with identically selected artificial samples extracted from the N-body simulations of Klypin et al. (1993). Our aim is to compare results in real and redshift space.

While, for simulated data, it is straightforward to work either in real or redshift space, from the observational point of view direct information on true galaxy distances (and thus on their peculiar velocities) is available only for restricted samples, typically limited to $\sim 6000\,\mathrm{km\,s^{-1}}$ and still affected by residual observational problems. Therefore, no direct information is available on the real space distribution for most of the galaxies in large redshift surveys like PPS, although a number of efforts are being carried on in that direction (see, e.g., Giovanelli & Haynes, 1994).

Statistical estimators are obviously affected by the fact of mapping the Universe in redshift space. In the case of the 2-point correlation function $\xi(r)$ significant distortions arise because of two main reasons. (i) On small scales, $\xi(r)$ is depressed because of the redshift-space elongation of high-velocity-dispersion regions (rich clusters). (ii) At large separations, $\xi(r)$ is amplified because of coherent motions inflowing towards overdense regions and outflowing from underdense regions (Kaiser 1987). These effects can be appreciated by analyzing the correlation function with respect to two components of the separation of galaxy pairs ($\theta$ and $\pi$, parallel and perpendicular to the line of sight, respectively; see, e.g., Davis & Peebles 1983, Fisher et al 1994). A similar decomposition can be performed on the power spectrum $P(k)$ (see, e.g., Cole, Fisher & Weinberg 1994). A number of investigations on redshift-space distortions for second-order statistics ($\xi(r)$ or $P(k)$) have been also performed using N-body simulations (see, e.g., Gramann, Cen & Bahcall 1993).

Redshift space distortions on the higher moments of galaxy distribution, have been recently addressed by Matsubara & Suto (1994), Suto & Matsubara (1994) and Bonometto et al (1994, hereafter BBGKP). The former two authors made use of CDM simulations only. In the latter paper the simulations considered here were already partially taken into considerations.

A previous analysis of galaxy data using neighbour moments (Bonometto et al. 1993, hereafter BIGGH) was able to exhibit deviations from a pure hierarchical model (HM) behaviour of 3–point and 4–point functions. In this work PPS was used after correcting the large 'fingers of God' by statistically collapsing each Abell cluster to its most probable spatial size. This operation had also allowed to recover the correct shape of $\xi(r)$ at small scales.

Our aim here is to verify the influence of using data in redshift space on the detection of deviations from a HM behaviour. We find that such features, when present in real space, are mostly erased by the passage to redshift space. As we shall discuss shortly, the passage to redshift space is a magnifier of nonlinear distortions; this will allow in turn to point out how peculiar velocities influence LS structure data.

## 2. PPS Results

The PPS database was compiled by Giovanelli and Haynes during the last decade (see Giovanelli & Haynes 1991). It consists mainly of highly accurate 21–cm HI line redshifts, partly unpublished, obtained with the NAIC 305–m telescope in Arecibo[1] and with the NRAO[2] 300–foot telescope formerly in Green Bank (Giovanelli, Haynes & Chincarini 1986; Haynes et al. 1988 ; Giovanelli and Haynes 1989). The radio data are complemented with optical observations of early–type galaxies carried out at the 2.4–m telescope of the McGraw–Hill Observatory [3] (Wegner, Haynes & Giovanelli 1993). In order to exclude regions of sky of high galactic extinction, we limit the sample to the region bound by $22^h \leq \alpha \leq 3^h\, 10^m$, $0° \leq \delta \leq 42°\, 30'$. We then select, after correcting Zwicky magnitudes ($m_{Zw}$) for extinction using the absorption maps of Burstein & Heiles (1978), all galaxies brighter than $m_{Zw} = 15.5$. The resulting sample is 98 % complete to this limiting magnitude, and includes 3395 galaxies.

---

[1] The Arecibo Observatory is part of the National Astronomy and Ionosphere Center, operated by Cornell University under a cooperative agreement with the National Science Foundation.

[2] NRAO: the National Radio Astronomy Observatory is operated by Associated Universities, Inc., under a cooperative agreement with the National Science Foundation.

In addition to the increased completeness, the sample used here differs from that used in BIGGH in no correction being applied to the observed velocities, apart from performing the transformation of Yahil, Tammann & Sandage (1977) to the centroid of the Local Group. As in BIGGH, our statistical analyses are performed on a volume–limited sample (VLS), by requiring that galaxies have $M \leq -19 + 5\log h$ and distance $d \leq 79\, h^{-1}$Mpc. In this new version, the VLS contains 1032 galaxies, 59 more than the VLS used in BIGGH.

Neighbour moments can be related to clustering models using suitable relations. A detailed treatment of this point, as well as a comparison between the use of cell and neighbour moments, is given in BBGKP. The analytical relations in use here are given in BIGGH (see also Peebles 1980, and references therein). The measures made on the samples amount to counting the numbers $N_k(r)$ of galaxies in spheres of radius $r$, centered around the $k$–th galaxy. Such numbers are compared with the numbers $n_k(r)$ of points of random samples occupying the same volume of VLS. From such numbers the quantities $C_{nm}(r) = \langle [N(r)]^n \rangle / \langle [n(r)]^m \rangle$ are estimated. From $C_{nm}(r)$ we work out the parameters of the so–called *hierarchical clustering* model (HM).

Besides of taking $\xi(r) = (r_o/r)^\gamma$, in HM the 3 and 4–point connected functions read

$$\xi^{(3)}_{123} = Q(\xi_{12}\xi_{13} + \xi_{21}\xi_{23} + \xi_{32}\xi_{31})$$

$$\xi^{(4)}_{1234} = Q_{4a}(\xi_{12}\xi_{13}\xi_{14} + (\text{symm. terms})) + Q_{4b}(\xi_{12}\xi_{23}\xi_{34} + (\text{symm. terms})) \,. \qquad (1)$$

The LS distribution fits HM if $Q$'s, worked out from $C_{nm}$ for different $r$'s, are compatible with a single value.

---

[3] The McGraw–Hill Observatory is located on Kitt Peak mountain, and jointly operated under a cooperative agreement by Dartmouth College, the Massachussets Institute of Technology and the University of Michigan.

An essential ingredient to our approach is the estimate of errors. This is carefully described in BIGGH and essentially amounts to an improved bootstrap procedure, which is repeated a large number of times in order to test the Gaussian distribution of errors.

We re-estimated the parameters of $\xi(r)$ on the VLS by applying the usual standard estimator described in BIGGH. This yielded $r_o = 8 \pm 1.2\,h^{-1}$Mpc and $\gamma = 1.24 \pm 0.11$ (bootstrap $3\sigma$ errors). In fig. 1 and fig. 2 the behaviour of $Q$ and $Q_4 \equiv Q_{4a}/3 + Q_{4b}$ as a function of $r$ is shown, calculated using the same algorithm of BIGGH, whose results are also plotted. Error bars correspond to $3\,\sigma$'s but, of course, points for different $r$ are not statistically independent. The present result – in pure redshift space – is fully consistent with HM, quite differently from that of BIGGH, obtained as previously discussed after a number of corrections for virial fingers and Virgo infall, in order to recover real positions from redshift space positions. The result shown in figs. 1 and 2 is one of the main outputs of this work.

It is clear that the scale dependence of $Q$'s, found in BIGGH, can be attributed either to actual features of the real-space distribution, or to excess artificial corrections introducing an apparent deviation from HM. A choice between these possibilities can be better performed after suitably analyzing the results of the N-body simulations.

## 3. Simulation Results in Real and Redshift Space

The final outputs from N-body simulations permit to build both real and redshift space artificial samples. Here, we use the PM simulations performed by Klypin et al. (1993), which have evolved $50\,h^{-1}$Mpc cubic boxes ($h$ is the Hubble constant in units of $100\,\mathrm{km\,s^{-1}Mpc^{-1}}$), with a formal resolution of $97.7\,h^{-1}$kpc, thanks to a $512^3$ set of grid points. Four different simulations are considered: two realizations of the Cold+Hot Dark Matter (CHDM) model, and a Cold Dark Matter (CDM) model with bias factor ($b$) 1.5 and 1. Further details can be found in BBGKP and in Klypin et al. (1994).

Since our aim is a comparison with a volume-limited sample of bright ($M \leq -19$) galaxies, let us describe in some detail the procedure followed to locate *galaxies* in the simulation box. Let $l$ be the side of the box and $n$ be the number density of VLS. $N^{(g)} = nl^3$ galaxies will be set in $N^{(p)}(\leq N^{(g)})$ peaks with mass $M_k \geq m$, such that $\sum_{k=1}^{N^{(p)}} \text{int}(m/M_k) = N^{(g)}$ ($M_k$ is obtained averaging over a 27 cells cube, whose side is $s = 293\,h^{-1}$kpc). The $\nu_k = \text{int}(m/M_k)$ galaxies of the $k$–th peak should be given a mass $m_k = M_k/\nu_k$; clearly $m \leq m_k < 2m$, but many peaks yield $\nu_k \gg 1$ and therefore most $m_k$'s are just slightly above $m$. Such mass distribution is not related to the actual mass function and is therefore replaced by an equal mass $\bar{m}$ for all galaxies. (We have also implemented an algorithm to distribute mass according to a Schechter law. This will be discussed elsewhere, as its use does not modify the conclusions of this work.)

Simulation outputs provide coordinates and velocities for each peak. However peak velocities are not sufficient to determine the position of the related galaxies in redshift space. In fact, virial equilibrium in the $k$–th peak prescribes that the $\nu_k$ galaxies it contains have a *rms* velocity $\bar{v}_k \simeq (4\pi/3)^{1/6} \nu_k^{1/2} (G\bar{m}/s)^{1/2} = [\nu_k h(\bar{m}/10^{10} M_\odot)]^{1/2} 86\,\text{km s}^{-1}$. With $\nu_k \sim 10$, the assumption of virial equilibrium creates an uncertainty in position of some Mpc's.

In fig. 3 we plot the overall distribution of velocities along a given line of sight, relative to peaks, obtained by attributing maxwellian velocities to galaxies (rms. velocity $\bar{v}_k$ within the $k$–th peak) in random directions. This shows how relevant are expected velocities inside peaks to induce distortions in redshift space. (Note the remarkable difference between CHDM and CDM's). Redshift space settings have been obtained by taking such velocities into account. We constructed several VLS's with shape and volume equal to the $M \leq -19$ PPS subsample. In implementing this procedure, there are some residual problems.

Among them, the most important point arises from the fact that the greatest depth obtainable from our $50 h^{-1}$Mpc box is $86.6 h^{-1}$Mpc. This exceeds the $79 h^{-1}$Mpc of the observational VLS, but such condition can be maintained only within a small solid angle around the diagonal of the box. Outside the box one has to consider replicae of the points of the simulation, owing to periodical boundary conditions. Not all replicated volumes however contain points already used and the line of sight is always differently directed in those zones which are used more than once. Multiple use occurs for $\sim 20\,\%$ of the box. Although the availability of a larger simulation box would be most welcome, here we are mostly concerned with the small–scale behaviour of clustering and velocities. Thus, for the present investigation it is certainly preferable the use of this kind of high resolution simulations, and there is no reason to suspect that the above large–scale limitations seriously affect our conclusions at $1$–$5h^{-1}$Mpc. This is supported by the comparison of results obtained from artificial VLS's with different observer positions, which weight differently the replicated regions. We find that our error estimates properly account also for the variations occuring when the observer is changed. Our plots are however obtained considering an observer at a fixed setting.

In fig. 4 we plot the hierarchical coefficient $Q$ of the 3–point function as a function of scale, in real and redshift space, for the four simulation outputs. In fig. 5 similar results are shown for the 4–point function. An effect close to the one detected in PPS is visible in both CHDM and in the $b = 1.5$ CDM simulations. The effect is not detected in the unbiased CDM model.

## 4. Discussion

The effect related to the passage from redshift to real space in simulations is not so prominent as the one due to the passage from redshift space to *corrected* VLS for PPS. The comparison with artificial VLS's therefore leaves us with a doubt whether the extra effect for PPS is due to excess *corrections* or to actual features of real data. However this does not affect the conclusion that, when a deviation from HM is present in real space, it tends to be erased in the passage to redshift space.

Fry & Gaztañaga (1994) had attempted to compare real and redshift space behaviours for various samples, within the count in cells approach. They were unable to detect deviations from HM. Virial finger compression, neighbour approach and the use of a deeper sample allow us to show such deviations for observational data.

That the passage from real to redshift space could damp deviations from HM has been more or less explicitly claimed elsewhere, by authors analysing simulation outputs. N–body simulations were studied to verify whether they support HM. However, in this context, Bouchet & Hernquist (1992) and Lahav et al. (1992) found deviations from HM at the passage from the linear to the non–linear regime. They considered this to be at variance in respect to redshift space observational data and ascribed the discrepancy to redshift space distortions, suggesting that HM parameters appear more constant in redshift space than in real space. In two recent papers Matsubara & Suto (1994) and Suto & Matsubara (1994) examined the effects of passing to redshift space on the basis of CDM N–body simulations. They also find that any signature of non–hierarchical behaviour in real space is erased by the passage to redshift space. Colombi et al. (1993) interpreted discrepancies from HM as due just to lack of statistics. BBGKP also showed in their N–body simulations that, although the distribution of dark matter particles exhibits marked differences when passing from real to redshift space, the distribution of galaxies is not affected as long as only peculiar motions of the peaks are taken into account.

Unlike previous papers, in this work we perform a parallel analysis of observational data and simulation outputs.

The shape of the dependence of $Q$'s on $r$ can be related to a clustering model different from HM. If one assumes that

$$\langle \delta N_1 ... \delta N_k \rangle = \delta V_1 ... \delta V_k n_V^k \prod_{i<j=1}^{k} (1 + \xi_{ij}) \qquad (2)$$

(Kirkwood superposition expression, hereafter KW), the connected 3–point function reads

$$\xi_{123}^{(3)} = Q(\xi_{12}\xi_{13} + \xi_{21}\xi_{23} + \xi_{32}\xi_{31}) + Q'\xi_{12}\xi_{23}\xi_{31} \qquad (3)$$

with $Q = Q' = 1$. In a similar fashion, the 4–point connected function contains terms up to the 6–th degree in $\xi$, with fixed coefficients. In BIGGH it was shown that a better fit to data is obtained with a KW–like expression, but with $Q = Q' \neq 1$ (similar results hold for the 4–point function).

Expressions of the kind shown in eq. (2) are obtained within the theory of biased galaxy formation (Jensen & Szalay 1986, Matarrese et al. 1986, Szalay 1988, Borgani & Bonometto 1990), assuming however that no displacement of galaxies took place from the site where they formed. As average galaxy velocities, almost independently of scale, range around 200–300 $h^{-1}$ km s$^{-1}$, KW features can be expected to be canceled below 1–2 $h^{-1}$Mpc and significantly attenuated up to 3–4 $h^{-1}$Mpc. Slightly above the latter limit we enter the linearity regime, where higher power terms become less and less important. Deviations from HM related to an initially biased galaxy formation, can therefore be expected to be still visible *in real space* in the range 2–5 $h^{-1}$Mpc, were they were actually detected. A countercheck of such result is the behaviour of the unbiased CDM simulation, where the KW–like regime seems to be absent, as expected.

Let us also outline that a detection of a KW–like regime is also due to the effectiveness of the neighbour approach to constrain the value of the constants $Q$ at large scales. Results over these scales are strictly compatible with outputs from the cell moment method, but in the latter case the number of independent cells at large separations is small, and error bars are obviously larger. The passage from real to redshift space does however act as a nonlinear mixing (i.e. peculiar velocity) magnifier: even the residual interval where primeval biased distribution had left an imprint, is compressed to nil.

Our tentative conclusion based on the results discussed here, is that the galaxy distribution does exhibit a physical dependence of the HM coefficient on scale. Such dependence is very similar to the one recently found by Gatzañaga (1994) by examining the APM angular data. This can be considered a confirmation of the success of the empirical virial finger compression procedure, applied to PPS. More generally, it seems fair to conclude that an analysis of non–Gaussian features of LS galaxy distribution is hard to perform in redshift space. Suitable improvements to empirical techniques of real space reconstruction should be carefully studied as an alternative to comparing data and simulations in redshift space.

ACKNOWLEDGMENTS. Thanks are due to Stefano Borgani and Angela Iovino for discussions. Angela Iovino also kindly allowed us to make use of an algorithm she had collaborated to build for a previous analysis. JRP acknowledges support from the NSF. AK and JRP utilized the CONVEX C3880 at the NAtional Center for Supercomputing Applications, University of Illinois at Urbana–Champaign.

Peebles P.J.E., 1980, The Large–Scale Structure of the Universe (Princeton: Princeton University Press)

Suto Y. & Matsubara, 1994, ApJ, 420, 504

Szalay, A.S. 1988, ApJ 333, 21

Yahil, A., Tammann, G.A., & Sandage, A., 1977, ApJ, 217, 903

Wegner, G., Haynes, M.P., & Giovanelli, R., 1993, AJ, 105, 1243
12

**Figure Captions**

Fig. 1 – The hierarchical coefficient $Q$ of the 3–point function for the $M \leq 19$, $d \leq 79 h^{-1}$ Mpc volume–limited sample (VLS) of the Perseus–Pisces Redshift Survey is shown as a function of the scale $r$. $Q$ is evaluated by the neighbour–count technique using spheres of radius $r$ and is expected to be a constant if the hierarchical model holds. Filled circles refer to the distribution of galaxies with distances obtained through redshifts (redshift space). Open circles refer to the sample where a number of corrections (for virial fingers and Virgo infall) have been applied in a tentative reconstruction of the real space distribution at small scales. Open circles have been slightly displaced to the right to avoid overlapping of error bars. The latter ones are estimated by a bootstrap–like method and correspond to 3 standard deviations.

Fig. 2 – The linear combination $Q_4 \equiv Q_{4a}/3 + Q_{4b}$ of the coefficients of the hierarchical 4–point function is plotted against the scale $r$ for the Perseus–Pisces VLS. Symbols and error bars are as in fig.1.

Fig. 3 – The distribution of the line–of–sight rms. velocities of galaxies, relative to peaks, estimated through the virial theorem is shown for the simulation outputs. Only those galaxies that have at least one companion inside the same peak are considered (solely in this case indeed a peculiar velocity relative to the peak rest frame is assigned to the galaxies). The histograms show the fraction of galaxies in bins of 10 Km/sec for the two runs of the Cold+Hot Dark Matter Model (CHDM$_1$, CHDM$_2$) and for Cold Dark Matter with bias $b = 1.5$ (CDM1.5) and unbiased (CDM1).

Fig. 4 – The hierarchical coefficient $Q$ at varying scale for the VLS's built up from the simulation outputs. Filled squares are for the real space analysis and open circles for the redshift space one. Errors are estimated in the same way as for the Perseus–Pisces VLS; here error bars correspond to one standard deviation. Filled squares have been slightly displaced to the right for clarity.

Fig. 5 – The dependance on scale of the coefficient $Q_4$ is shown for the four artificial VLS's. Symbols and errors are as in fig.4.

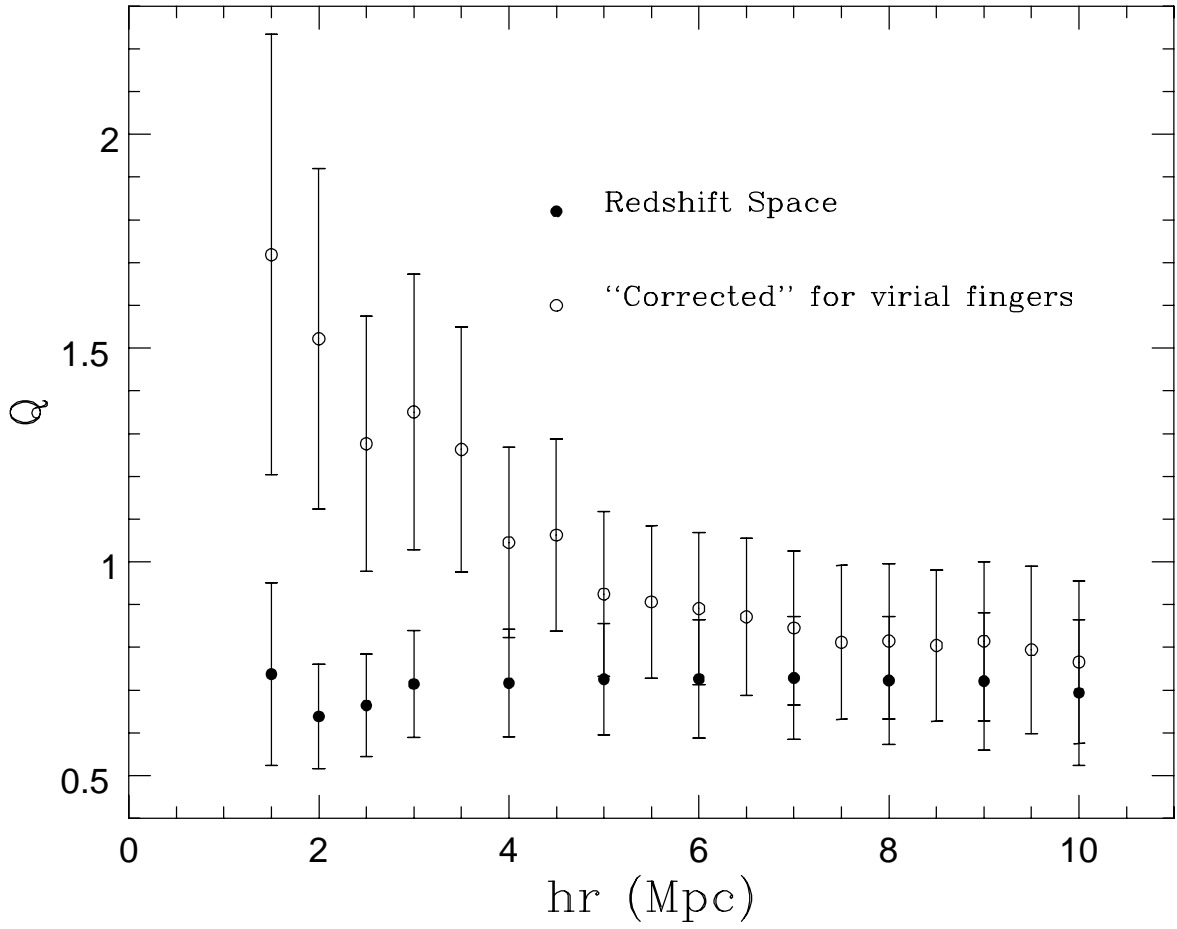

Figure 1

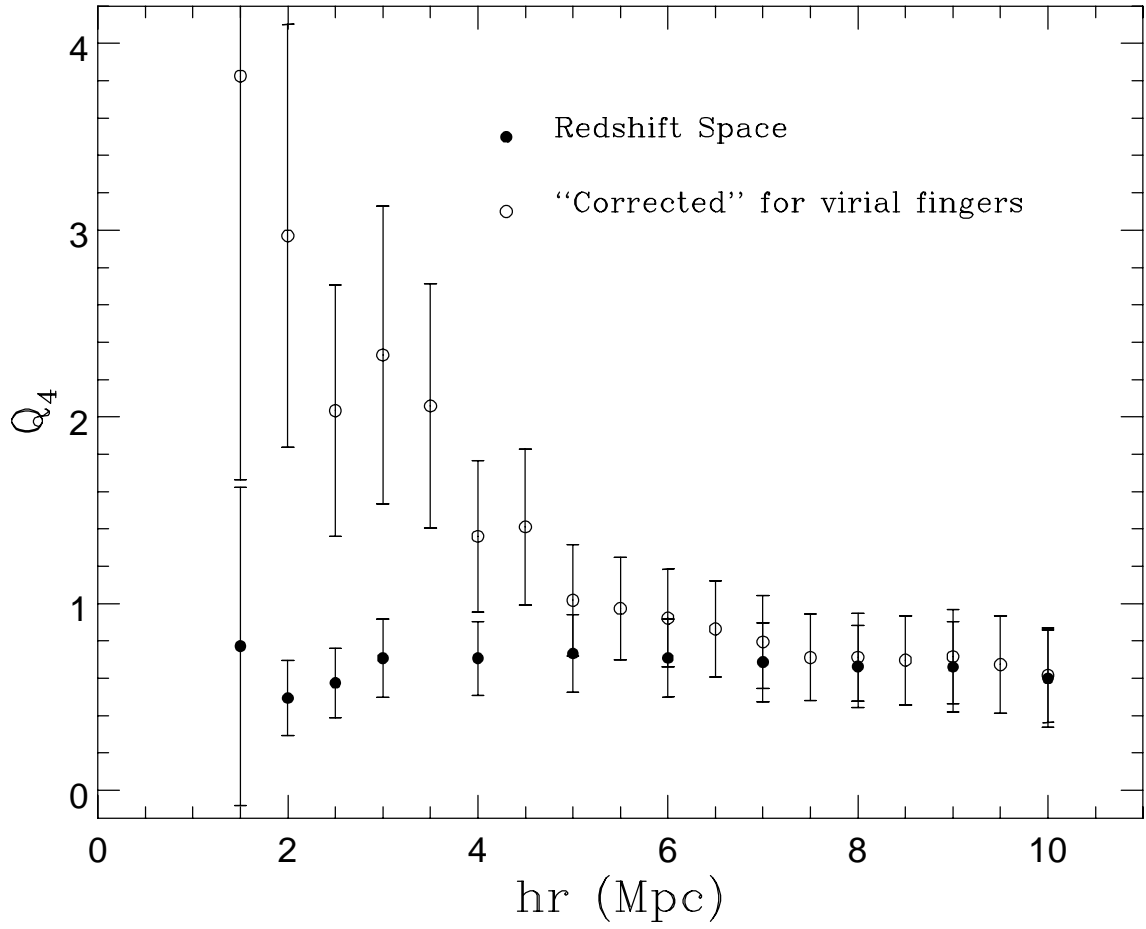

Figure 2

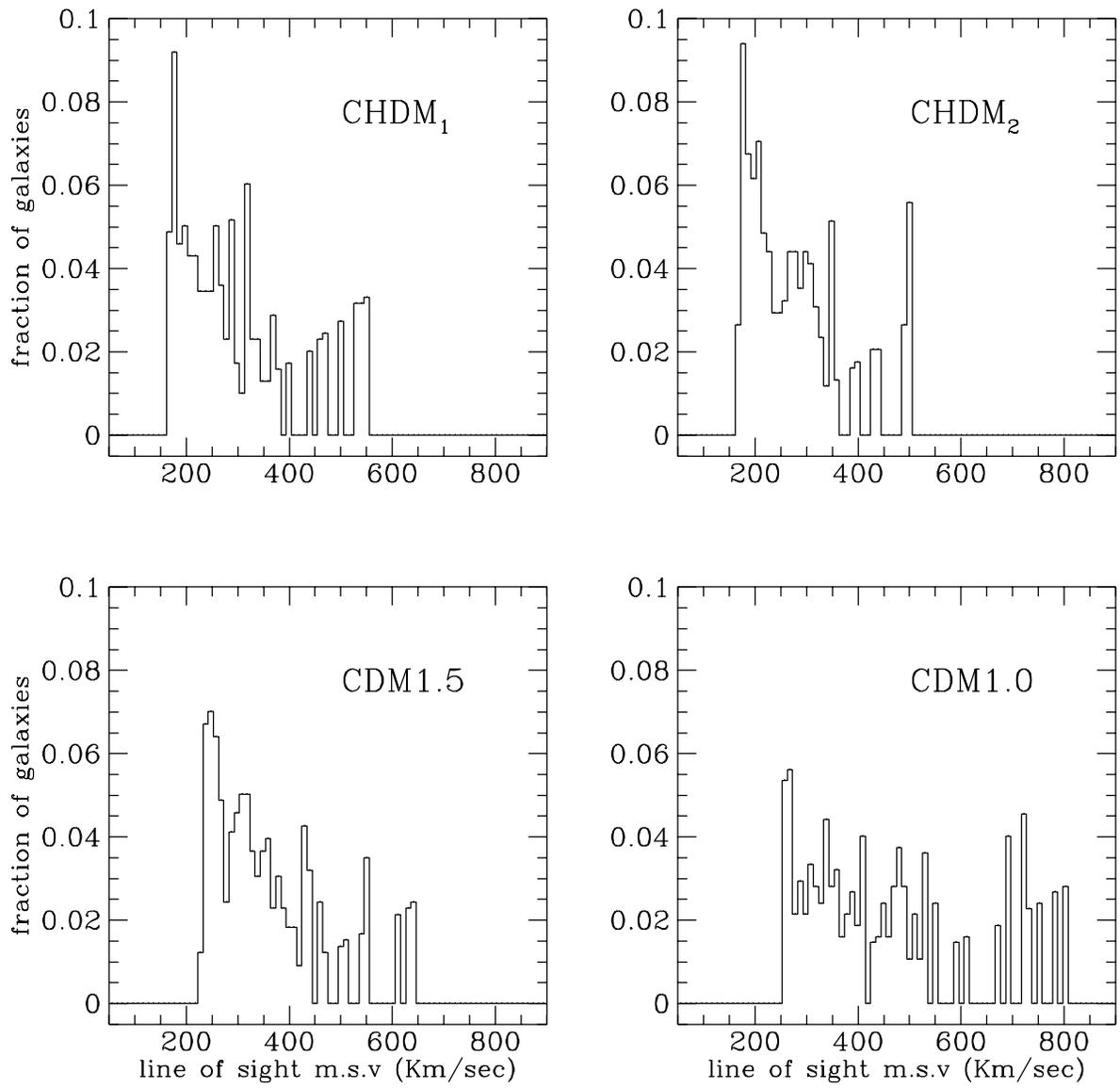

Figure 3

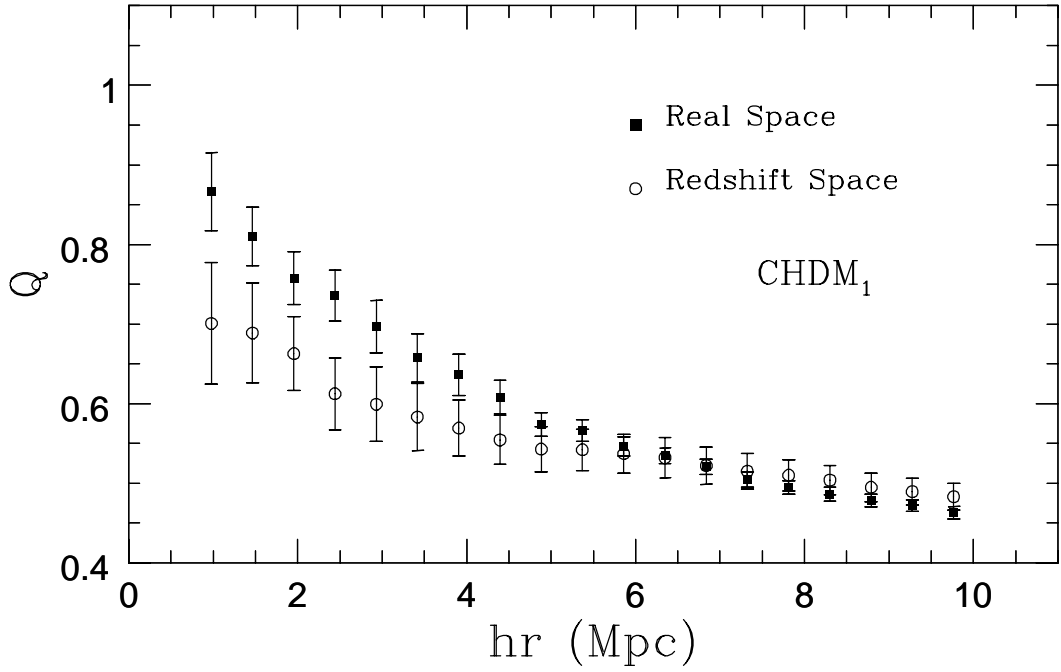

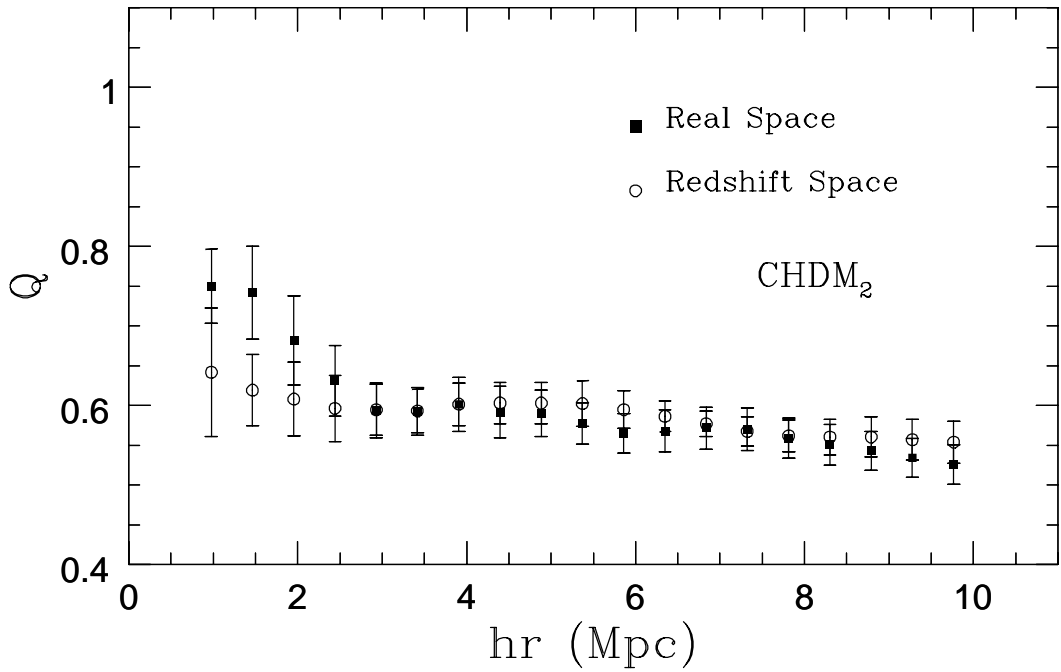

Figure 4

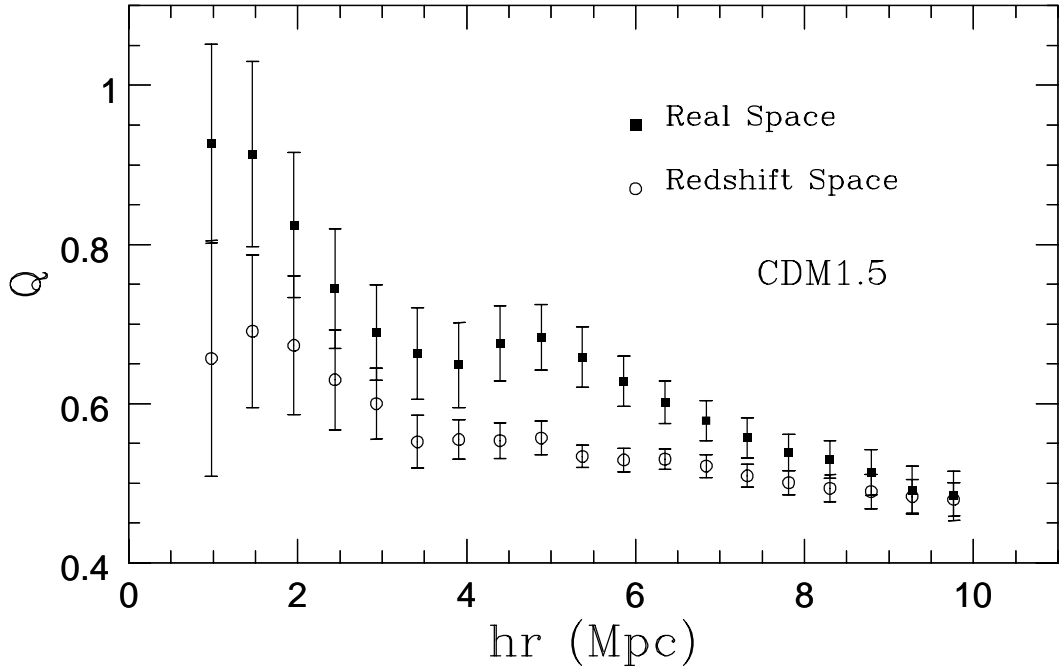

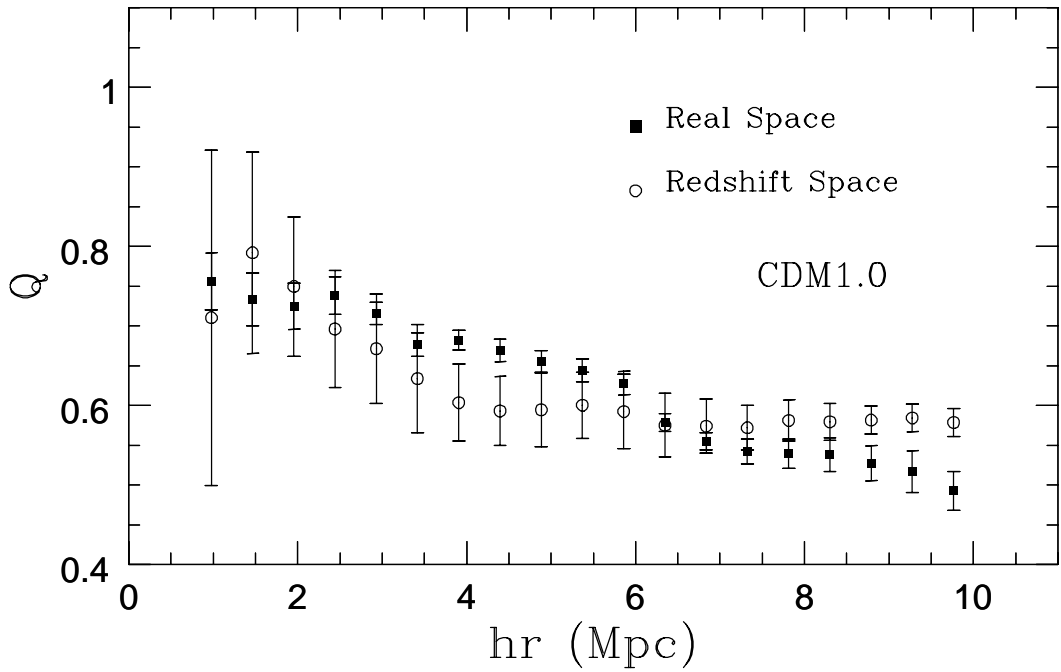

Figure 4

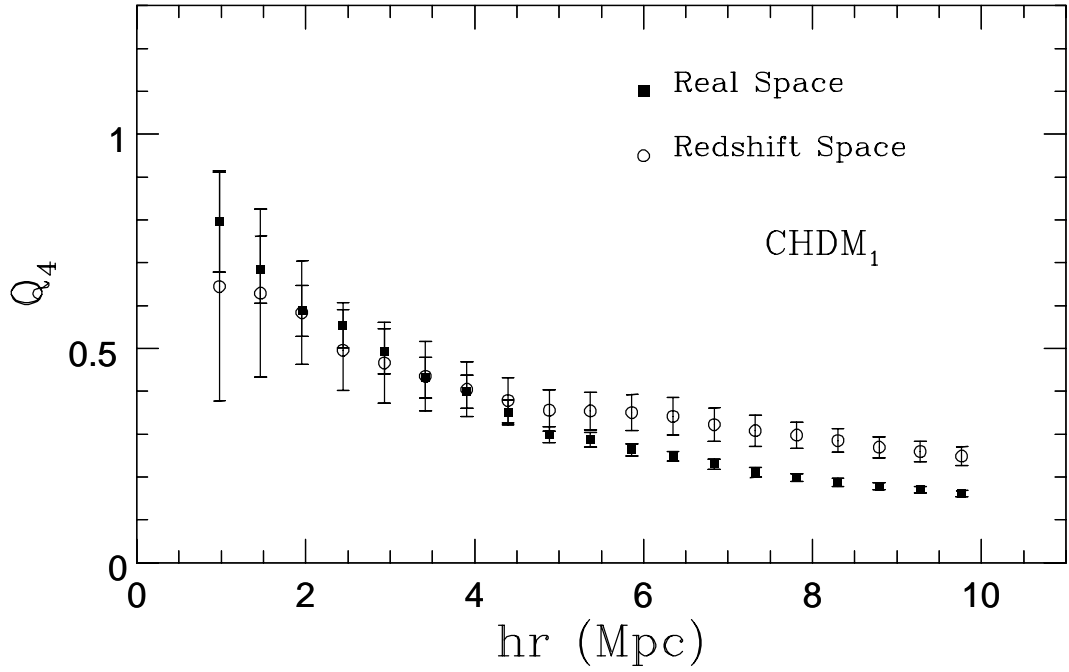
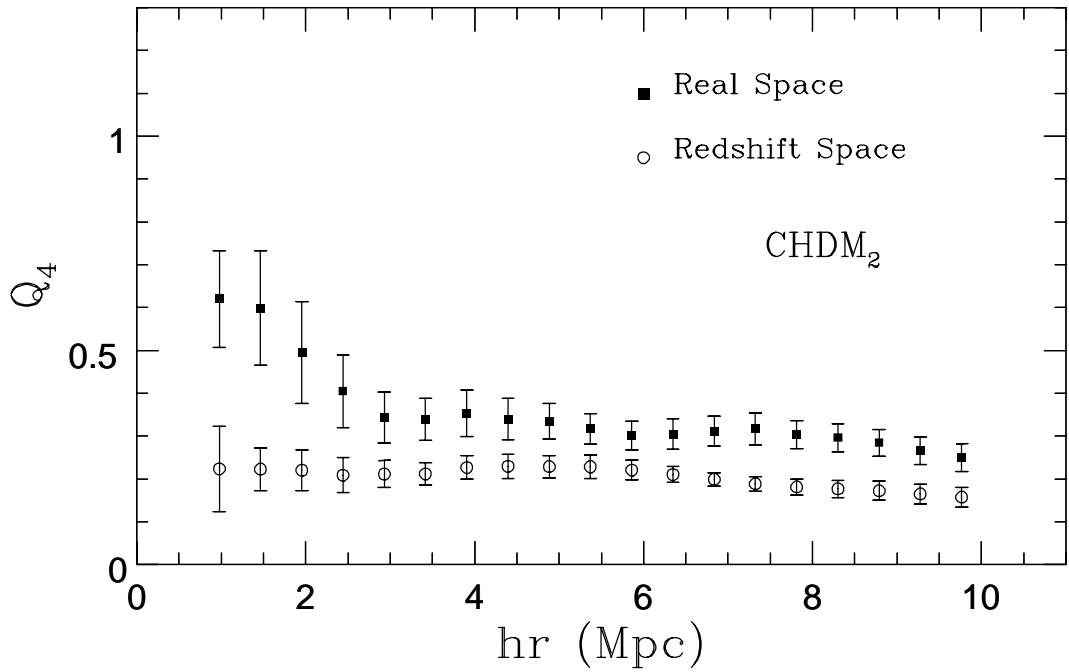

Figure 5

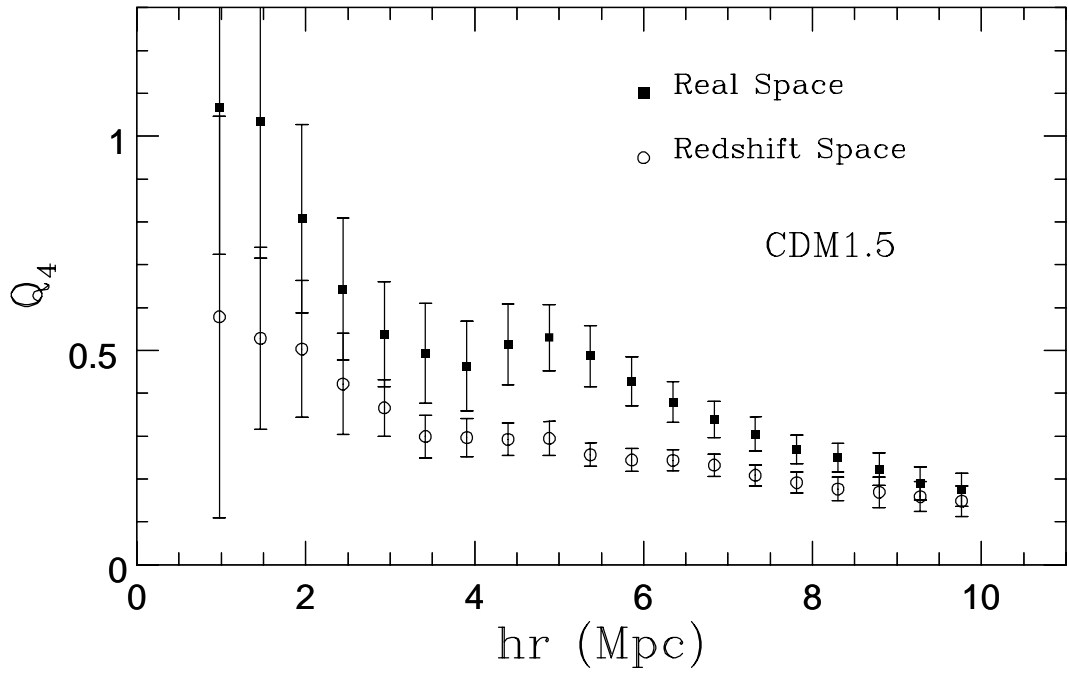

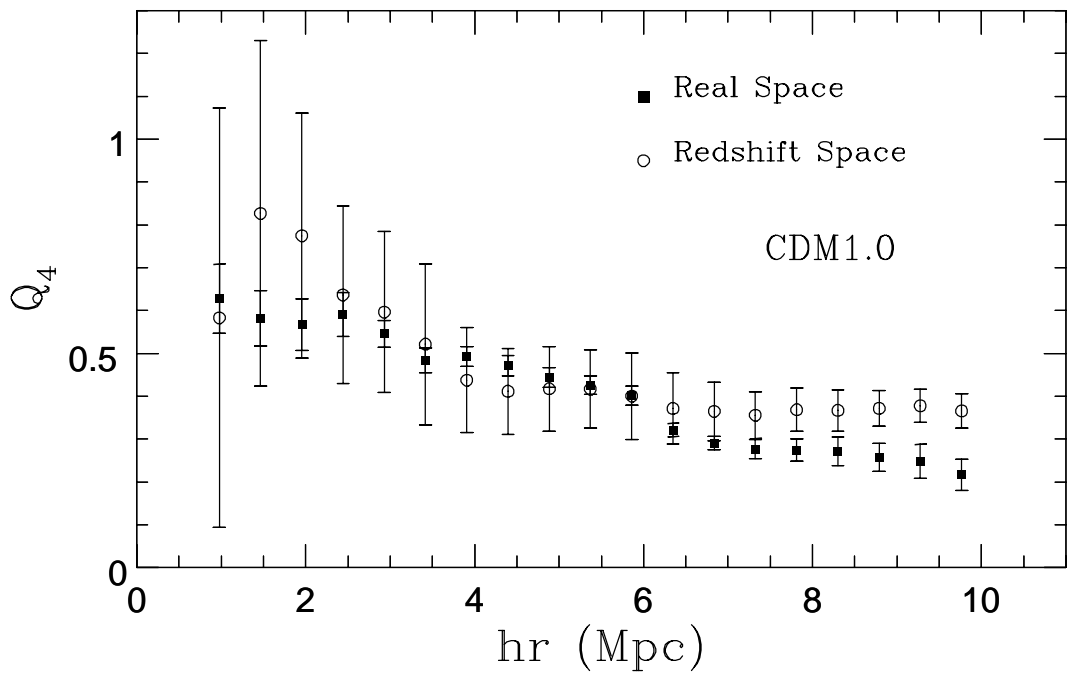

Figure 5